\documentclass[12pt]{article}
\usepackage{times}
\usepackage{geometry}
\usepackage[utf8]{inputenc}
\usepackage{enumitem,amssymb}
\usepackage{ragged2e}
\newlist{thematic}{itemize}{8}
\setlist[thematic]{label=$\square$}
\usepackage{pifont}
\usepackage{tabu}
\usepackage{multirow}
\usepackage{longtable}
\usepackage{rotating}
\usepackage{natbib}
\newcommand{\cmark}{\ding{51}}%
\newcommand{\done}{\rlap{$\square$}{\raisebox{2pt}{\large\hspace{1pt}\cmark}}%
\hspace{-2.5pt}}

\newcommand{\mnras}{MNRAS~}
\newcommand{\jcap}{JCAP~}
\newcommand{\aap}{A\&A}%
\newcommand{\apjl}{ApJL~}
\newcommand{\apjs}{ApJS~}
\newcommand{\apj}{ApJ~}
\newcommand{\aj}{AJ~}

\newcommand{\prd}{Phys.~Rev.~D}

\begin{document}
\onecolumn

\raggedright
\huge
Astro2020 Science White Paper \linebreak

The Next Decade \\
of Astroinformatics and Astrostatistics \linebreak

\normalsize

\noindent \textbf{Thematic Areas:} \hspace*{60pt} $\done$ Planetary Systems \hspace*{10pt} $\done$ Star and Planet Formation \hspace*{20pt}\linebreak
$\done$ Formation and Evolution of Compact Objects \hspace*{31pt} $\done$ Cosmology and Fundamental Physics \linebreak
  $\done$  Stars and Stellar Evolution \hspace*{1pt} $\done$ Resolved Stellar Populations and their Environments \hspace*{40pt} \linebreak
  $\done$    Galaxy Evolution   \hspace*{45pt} $\done$             Multi-Messenger Astronomy and Astrophysics \hspace*{65pt} \linebreak
  
\textbf{Principal Author:}

Name: {\it Aneta Siemiginowska}	
 \linebreak						
Institution:  Center for Astrophysics $|$ Harvard \& Smithsonian \\
    {\it Chair, AAS Working Group on Astroinformatics and Astrostatistics}
 \linebreak
Email: asiemiginowska@cfa.harvard.edu
 \linebreak
 \linebreak

\textbf{Co-authors:} 

Gwendolyn Eadie\footnotemark[1]\footnotemark[2]\footnotemark[3],
Ian Czekala\footnotemark[4],
  Eric Feigelson\footnotemark[5],
  Eric B. Ford\footnotemark[5],
  Vinay Kashyap\footnotemark[6],
  Michael Kuhn \footnotemark[7],
  Tom Loredo\footnotemark[8],
  Michelle Ntampaka\footnotemark[9],
  Abbie Stevens\footnotemark[10]$^,$\footnotemark[11],
  Arturo Avelino\footnotemark[6],
  Kirk Borne\footnotemark[12],
  Tamas Budavari\footnotemark[13],
  Blakesley Burkhart\footnotemark[6],
  Jessi Cisewski-Kehe\footnotemark[14],
  Francesca Civano\footnotemark[6],
  Igor Chilingarian\footnotemark[6],
  David A. van Dyk\footnotemark[15],
  Giuseppina Fabbiano\footnotemark[6],
  Douglas P. Finkbeiner\footnotemark[9],
  Daniel Foreman-Mackey\footnotemark[16],
  Peter Freeman\footnotemark[17],
  Antonella Fruscione\footnotemark[6],
  Alyssa A. Goodman\footnotemark[9],
  Matthew Graham\footnotemark[7],
  Hans Moritz Guenther\footnotemark[18],
  Jon Hakkila\footnotemark[19],
  Lars Hernquist\footnotemark[9],
  Daniela Huppenkothen\footnotemark[2]$^,$\footnotemark[3],
  David J. James\footnotemark[6],
  Casey Law\footnotemark[4],
  Joseph Lazio\footnotemark[20],
  Thomas Lee\footnotemark[21],
  Mercedes López-Morales\footnotemark[6],
  Ashish A. Mahabal\footnotemark[22],
  Kaisey Mandel\footnotemark[23],
  Xiao-Li Meng\footnotemark[9],
  John Moustakas\footnotemark[24],
  Demitri Muna\footnotemark[25],
  J. E. G. Peek\footnotemark[26]$^,$\footnotemark[27],
  Gordon Richards\footnotemark[28],
  Stephen K.N. Portillo\footnotemark[2]$^,$\footnotemark[3],
  Jeff Scargle\footnotemark[29],
  Rafael S. de Souza\footnotemark[30],
  Joshua S. Speagle\footnotemark[9],
  Keivan G. Stassun\footnotemark[31],
  David C. Stenning\footnotemark[15],
  Stephen R. Taylor\footnotemark[22],
  Grant R. Tremblay\footnotemark[6],
  Virginia Trimble\footnotemark[32],
  Padma A. Yanamandra-Fisher\footnotemark[33],
  C. Alex Young\footnotemark[34].

\hfill

\begin{raggedright}

\footnotetext[1]{\small{eScience Institute, University of Washington, Seattle, WA 98195, USA}}
\footnotetext[2]{\small {DIRAC Institute, Department of Astronomy, University of Washington, Seattle, WA 98195, USA}}
\footnotetext[3]{\small{Department of Astronomy, University of Washington, Seattle, WA 98195, USA}}
\footnotetext[4]{\small{Department of Astronomy, University of California, Berkeley, CA 94720  USA}}
\footnotetext[5]{\small{Penn State University, University Park, PA 16802, USA}}
\footnotetext[6]{\small{Center for Astrophysics $|$ Harvard \& Smithsonian, Cambridge, MA 02138, USA}}
\footnotetext[7]{\small California Institute of Technology, Pasadena, CA 91109, USA}
\footnotetext[8]{\small{Cornell University, Cornell Center for Astrophysics and Planetary Science (CCAPS) \& Department of Statistical Sciences,  Ithaca, NY 14853, USA}}
\footnotetext[9]{\small{Harvard University, Cambridge, MA 02138, USA}}
\footnotetext[10]{\small{Department of Physics \& Astronomy, Michigan State University, East Lansing, MI 48824, USA}}
\footnotetext[11]{\small{Department of Astronomy, University of Michigan, Ann Arbor, MI 48109, USA}}
\footnotetext[12]{\small{Booz Allen Hamilton, Annapolis Junction, MD, USA}}
\footnotetext[13]{\small{Department of Applied Mathematics \& Statistics, Johns Hopkins University, Baltimore, MD 21218, USA}}
\footnotetext[14]{\small{Department of Statistics \& Data Science, Yale University, New Haven, CT 06511, USA}}
\footnotetext[15]{\small {Department of Mathematics, Imperial College London, SW7 2AZ, UK}}
\footnotetext[16]{\small{Flatiron Institute, Center for Computational Astrophysics, New York, NY 10010}}
\footnotetext[17]{\small {Carnegie Mellon University, Pittsburgh, PA, USA}}
\footnotetext[18]{\small{Massachusetts Institute of Technology,
Kavli Institute for Astrophysics and Space Research, Cambridge, MA 02139, USA}}
\footnotetext[19]{\small {Department of Physics \& Astronomy, Associate Dean of the Graduate School, University of Charleston, Charleston, SC 29424, USA}}
\footnotetext[20]{\small{Jet Propulsion Laboratory, California Institute of Technology, Pasadena, CA 91109, USA}}
\footnotetext[21]{\small {University of California Davis, CA 95616, USA}}
\footnotetext[22]{\small{TAPIR Group, Division of Physics, Mathematics, \& Astronomy, California Institute of Technology,  Pasadena, CA 91125, USA}}
\footnotetext[23]{\small {University of Cambridge, Cambridge, CB3 0HA, UK}}
\footnotetext[24]{\small{Department of Physics \& Astronomy, Siena College,
Loudonville, NY 12211, USA}}
\footnotetext[25]{\small{Center for Cosmology and AstroParticle Physics, The Ohio State University, Columbus, OH 43210, USA}}
\footnotetext[26]{\small{Department of Physics \& Astronomy, Johns Hopkins University, Baltimore, MD 21218, USA}}
\footnotetext[27]{\small{Space Telescope Science Institute, Baltimore, MD 21218, USA}}
\footnotetext[28]{\small {Drexel University, Department of Physics, Philadelphia, PA 19104}}
\footnotetext[29]{\small {Space Science Division, NASA Ames Research Center,
Moffett Field, CA 94035-0001}}
\footnotetext[30]{\small{Department of Physics \& Astronomy, University of North Carolina at Chapel Hill, Chapel Hill, NC 27599, USA}}
\footnotetext[31]{\small {Vanderbilt School of Engineering, Vanderbilt University,
Nashville, TN 37235, USA}}
\footnotetext[32]{\small University of California, Irvine, CA 92697, USA}
\footnotetext[33]{\small{Space Science Institute, Boulder, CO 80301, USA}}
\footnotetext[34]{\small{NASA Goddard Space Flight Center, Greenbelt, MD 20771
USA}}
\end{raggedright}

\pagebreak
\newpage

\textbf{Abstract:}
Over the past century, major advances in astronomy and astrophysics
have been largely driven by improvements in instrumentation and data
collection. With the amassing of high quality data from new
telescopes, and especially with the advent of deep and large
astronomical surveys, it is becoming clear that future advances will
also rely heavily on how those data are analyzed and interpreted.  New
methodologies derived from advances in statistics, computer science,
and machine learning are beginning to be employed in sophisticated
investigations that are not only bringing forth new discoveries, but
are placing them on a solid footing. Progress in wide-field sky surveys, interferometric imaging, precision cosmology, exoplanet detection and characterization, and many subfields of stellar, Galactic and extragalactic astronomy, has resulted in complex data analysis challenges that must be solved to perform scientific inference. Research in astrostatistics and astroinformatics will be necessary to develop the state-of-the-art methodology needed in astronomy. Overcoming these challenges requires dedicated, interdisciplinary research. We recommend: (1) increasing funding for interdisciplinary projects in astrostatistics and astroinformatics; (2) dedicating space and time at conferences for interdisciplinary research and promotion; (3) developing sustainable funding for long-term astrostatisics appointments; and (4) funding infrastructure development for data archives and archive support, state-of-the-art algorithms,  and efficient computing.

\newpage

{\bf 1. What is the role of astrostatistics and astroinformatics research?}

$\rightarrow$ \underline{{\it To develop modern methods for extracting scientific information from astronomical data.}}

\vspace{1ex}

{\bf Astrostatistics} forms the foundation for robust algorithms and principled methods that are applied to a variety of problems in astronomy. 
{\bf Astroinformatics} involves the systematic and disciplined development of code, data management and dissemination techniques, high-performance computing, and machine learning based inference. Both astrostatistics and astroinformatics (i.e., astro data science) have been rapidly emerging {\bf fields of research} rigorously pursued at the intersection of observational astronomy, statistics, algorithm development, and data science \cite{borne2010, loredo2012, feigelson2013, vandyk2015,bigdata}. 
The number of articles with keyword `Methods: Statistical' increased by a factor of 2.5 in the past decade; those with `machine learning' increased by 4 times over five years; and those with `deep learning' have more than tripled every year since 2015. Thus, the challenges of astronomical sciences reveal a deep and broad demand for advanced methodology and techniques. {\it Astronomical problems impossible to approach with traditional methods are now forefront research efforts because 
of advancements in astrostatistics and astroinformatics.}

In the next decade, astronomy data will present new challenges, and will make 
astrostatistics and astroinformatics research a necessity for nontrivial scientific inference in an increasing range of critical research areas. Astronomy `big data' described by the four V's --- volume, velocity, variety, and veracity --- demand new methodologies. It is vitally important that the quality and sophistication of the techniques match the quality and sophistication of the data. The specific application of any new method requires research involving data, statistics, algorithm development and computations, and, thus, the combined knowledge and experience of astronomers, statisticians, and computational experts. Cross-disciplinary collaboration and communication at a very high level are critical to such research; conceptual and jargon barriers between disciplines must be overcome.

Several white papers on astrostatistics and astroinformatics research, endorsed by dozens of leaders in the fields, were submitted to the Astro2010 Decadal Survey 
\cite{loredo2009, borne2009a, borne2009b, ferguson2009}. Since then, some recommendations have been implemented, such as the formation of the Working Group on Astroinformatics \& Astrostatistics within the American Astronomical Society, and the Astrostatistics Interest Group within the American Statistical Association. {\it What remains underdeveloped, however, is the formal recognition of and financial commitment to the efforts needed to make necessary progress in astrostatistics and astroinformatics.}

Astrostatistics and astroinformatics research impacts all areas of astronomy and needs to be recognized as a science area within astronomy. explorations. 
Our recommendation is to: (1) create supportive environments for long-term research in astrostatistics and astroinformatics; (2) promote research in this field with specific national level programs, fellowships, professional development, and consulting; and (3) provide sustained funding for long-term research programs. 

\medskip

{\bf 2. How do modern astrostatistics and astroinformatics methods impact astronomy?}

$\rightarrow$ \underline{{\it They overcome challenges with data and improve scientific inference.}}

Astrostatistics and astroinformatics research does not fit traditional thematic boundaries, as it includes both technological development and scientific research in statistical and information sciences. However, 
these disciplines are 
now a necessity for modern astronomical research.
Tables 1 and 2 highlight recent advances and expected challenges, and indicate the impact of emerging methods in thematic areas of astronomy.

\bigskip

{\bf 3. How can the state-of-the-art methods be best applied 
in astronomy?}

$\rightarrow$ \underline{{\it Through astronomy involvement in active methodology research}}

\vspace{1ex}

Existing statistical and machine learning methods need to be further developed to be applicable in astronomy. For example, adaptation of recent machine learning advancements to address building {\it explanatory} models rather than task-specific {\it predictive} models requires astronomy involvement in two active research areas of machine learning:

{\it Scalable probabilistic machine learning (including deep
learning)}:
Most ML algorithms seek to make one set of predictions or
point estimates, optimal to one specific end task. In astronomy, methods need to
{\it quantify uncertainty} and provide results (e.g., probabilistic catalogs) that
enable {\it uncertainty propagation}.
Probabilistic methods 
are well suited to this, but are computationally expensive and not easily scalable to large datasets.
Scalable approaches using ML 
are being investigated.
Collaboration with statisticians and computer scientists is needed to
develop such methods tailored to astronomers' needs. {\bf Astronomy needs to become a {\it driver} of this research.}\\
 
\vspace{1ex}

{\it Interpretable machine learning (especially deep learning)}:
Complex machine learning methods are coming to astronomy (e.g., deep learning methods involving convolutional and adversarial nets for analyzing image data \cite{gans2018, fussell2018, pasquet2018, ntampaka2018}, and recurrent neural nets for time series data \cite{narayan2019}).  Unfortunately, such methods often are used as `black box' predictors, while generalizable understanding of a phenomenon requires an {\it interpretable} model.  The emerging field of {\it interpretable machine learning} involves explanatory goals, not just predictive goals. {\bf Astronomy needs
to {\it actively participate} in this research.}

\bigskip

{\bf 4. Recommendations}

The compilations in Tables 1 and 2 highlight two important facts: (1) common methodology is repeatedly used with small alterations across different wave-bands to address diverse problems that span many thematic areas; and
(2) duplicated development efforts slow the pace of advance.  
To facilitate the faster development and dissemination of advanced methods we recommend:

{\bf Funding:} 
Astrostatistics and astroinformatics must be recognized 
as a subfield of astronomical research that affects all of its thematic areas. Proposals in this field must be evaluated by appropriately cross-disciplinary panels.

{\bf Communication:} 
Astronomy conferences must make room for methodological discussion, both to disseminate new advances and to raise the awareness for non-experts.  Funding for tutorials and other means of communication should be encouraged.

{\bf Sustainability:} There must be sustained funding 
 through grants and fellowships, 
to support graduate students and post-docs for several years.  Astronomy departments should be encouraged to have more tenure-track positions focused on data science research.

{\bf Infrastructure:} 
There must be support for  
both maintaining data archives and training data sets, 
for publicly available and supported software, and efficient computing.

\begin{sidewaystable}
\small{
		    \centering
\leftline{Table 1: Science, Methodology, and Issues}
			\begin{tabu} to 1.05\textwidth{X[1,1]X[1,1]X[1,1]X[1,1]}
				\hline
{\bf Science Measurements}  & {\bf Traditional Methods}  & {\bf Limitations \& Challenges} & {\bf Emergent Methodologies } \\		
				\hline
				\noalign{\vskip 1mm}	
\multicolumn{4}{l}{\bf Distance Measurements } \\
\noalign{\vskip 1.5mm}
				e.g., to stars, dust, and quasars & inverting parallaxes, sample truncation, astrometry-based luminosity, template fitting, stellar variability, photo-z & biases, need bias corrections, uncertainties ignored & Bayesian Inference for distances from parallaxes and for proper motions from astrometric data \cite{luri2018}, machine learning methods,  photometric redshfits \cite{cavuoti2015,Elliott2015,almosallam2016, Beck2017, dejong2017,Salvato2018} \\
				\noalign{\vskip 1mm}
				\hline
				\noalign{\vskip 1mm}
				\multicolumn{4}{l}{\bf Mass Estimates} \\
				\noalign{\vskip 1mm}
				e.g., of the Milky Way, dwarf galaxies, supermassive black holes, galaxy groups and clusters & kinematic tracers, timing argument, hyper velocity stars, reverberation, mass-$\sigma$ relation, power-law scaling relations   & data incompleteness, large uncertainties, scatter, extrapolation to larger distances, over-simplified models, biases & Bayesian hierarchical models, Approximate Bayesian Computation (ABC), carma models \cite{kelly2009,kelly2011,kelly2013}; machine learning, Bayesian model averaging \citep{mcmillan2011, eadie2017, patel2017} \\
\noalign{\vskip 1mm}
				\hline
				\noalign{\vskip 1mm}
				\multicolumn{4}{l}{\bf Stellar Properties \& Evolution}\\
				\noalign{\vskip 1mm}
				e.g., stellar type, temperature, composition, metallicity, coronal composition, density, stellar evolution; & forward fitting physics-based models (with chemical networks, MHD and planets for protoplanetary disks); stellar evolution models, spectral lines fitting, isochrone fitting, catalog matching and membership classification & degenerate models and parameters,
				difficulty in model selection, uncertainty quantification, correlated measurement uncertainties; inefficient sampling methods (e.g., MCMC) and need simplifications to the forward fitting models & Gaussian Processes \cite{czekala2015b,dfm2017,dfm2018}, machine learning methods \cite{ting2018}, Bayesian inference; model independent data-driven approaches for rotation curves, matched-filter for line searches \\
				\noalign{\vskip 1mm}
				\hline
				\noalign{\vskip 1mm}
				\multicolumn{4}{l}{\bf Population Studies} \\
				\noalign{\vskip 1mm}
				e.g. source detection, structures of diffuse regions, classifying galaxies and stars; identifying moving groups, stellar populations, globular cluster populations & spectral line fitting, photometry, colour-magnitude diagrams; two-point correlation function \cite{Peebles80}  & overlapping sources, faint structures \cite{stein2015,mckeough2016}, non-Gaussian uncertainties, unknown populations, complex morphology \cite{CartwrightWhitworth04,Grasha19} & probabilistic catalogues \cite{brewer2013, jones2015, portillo2017, portillo2019}; machine learning for open clusters \cite{Cantat-Gaudin2018,regier2018}; identifying members of stellar groups \cite{gagne2018,Buckner19}, spatial point patterns \cite{Baddeley15}; account for uncertainties \cite{xu2014}; wavelet-based clustering methods \\
				\noalign{\vskip 1mm}
\hline
\end{tabu}
}
\end{sidewaystable}

\begin{sidewaystable}
\small{
    \centering
\begin{center}
\leftline{Table 2: Astroinformatics and Astrostatistics Research in Thematic Areas}
\begin{tabu} to 1.05\textwidth{X[1,l]X[1,l]X[1,l]}
\hline
{\bf Advances} (issues under consideration)     & {\bf (Data/Analysis) Challenges}   &   {\bf Future } (Emergent Methods/Methodology) \\
\hline
\noalign{\vskip 1.7mm}
\multicolumn{3}{l}{\bf Stars and Stellar Evolution: Magnetic Activity, Populations Evolution, Environment}  \\
\noalign{\vskip 1mm} 
Stellar cluster catalog matching and membership classification \cite{Broos+11,budavari2017};  structure in diffuse X-ray background \cite{Facundo+19}; solar feature classification and properties \cite{HEK2010,Stenning+2013}; flare modeling, energy release and evolution \cite{Aschwanden+2016}; thermal segmentation of the corona \cite{Stein+16}; stellar coronal thermal and density structure via Emission Measure distributions \cite{DEM98,Wood+2018}; sources of coronal heating (e.g., nanoflares) \cite{Cargill1994,Cranmer+2007}; effect of stellar activity on exoplanets. \cite{Cuntz.Shkolnik2002,Shkolnik+2013}
& Solar and stellar flare onsets and distributions \cite{Kashyap+2002}; characterization of stellar activity to reveal hidden signals of exoplanets \cite{Rajpaul+2015,nelson2014}; determining the nature (magnetic or tidal) and magnitude of the Star-Planet Interaction effect \cite{Cuntz+2000,Poppenhaeger+2013}; isochrone fitting to determine ages, metallicity, and star formation history of star clusters \cite{Bitsakis+2017,BAGPIPES2018}; completeness and limitations of the Heliophysics Event Knowledgebase \cite{McCauley+2015,Aggarwal+2018}
& 
Solar dispersed image spectral decomposition (also applicable to SNRs) \cite{Winebarger+2018}; solar and stellar DEMs that incorporate atomic data uncertainties \cite{Yu+2018}; disambiguating photons from overlapping close binaries in confused fields to facilitate spectral and timing analysis \cite{jones2015}; loop recognition in solar coronal images \cite{Aschwanden2010,Stenning+2013}; morphological analysis to recognize diffuse structure \cite{Picquenot+19,Fan+19}\\
\noalign{\vskip 1mm}
\hline
\noalign{\vskip 1.5mm} 
\multicolumn{3}{l}{\bf Formation and Evolution of Compact Objects} \\
\noalign{\vskip 1mm} 
X-ray spectral-timing analysis \cite{stevens2016}; Bayesian Inference and evolutionary algorithms for neutron star equation of state \cite{ozel2015,ozel2016,steiner2018}; merging systems\cite{roulet2019}; transient detections\cite{cabrera-vives2017}; accretion states \cite{uttley2011}.  &
Use of spectral, spatial, time and polarimetry domain \cite{rosa2019,ray2019}; periodicity detection \cite{vaughan2016}; "needle in a haystack" searches \cite{law2018a,law2018b}; state transitions \cite{uttley2014,heil2015,uttley2017}; localization \cite{corley2019} &
New models, computational power; Gaussian processes in time domain for Poisson (Cox process) X-ray and gamma-ray \cite{kelly2011,kelly2013,kelly2014,sobolewska2014}; use of higher order Fourier product and non-linear signal processing \cite{huppenkothen2018}; machine learning methods \cite{huppenkothen2017,mahabal2017,sedaghat2018,mahabal2019}
\\
\noalign{\vskip 1mm}
\hline
\noalign{\vskip 1.5mm} 
\multicolumn{3}{l}{\bf Galactic Astronomy and Galaxy Evolution} \\
\noalign{\vskip 1mm} 
Gaia data\cite{gaia2018,hogg2018};  Machine Learning methods and Bayesian inference  & Account for uncertainties, incompleteness and biases & Machine learning to discover new stellar open clusters \cite{Cantat-Gaudin2018}; photometric redshifts \cite{cavuoti2015,Elliott2015,almosallam2016, Beck2017, dejong2017,Salvato2018}; Bayesian Inference for distances from parallaxes and for proper motions from astrometric data \cite{hogg2018a,luri2018}; the mass of the Milky Way  \cite{mcmillan2011, eadie2017, patel2017, callingham2019}, identifying members of stellar groups \cite{gagne2018} \\
\noalign{\vskip 1mm} 
\hline
\noalign{\vskip 1mm} 
\multicolumn{3}{l}{\bf Multi-Messenger Astronomy and Astrophysics}\\
\noalign{\vskip 1mm} 
Detecting transients, multi-band identifications & Localization,  nanohertz GW detection \cite{law2018a, law2018b}, GW-EM coincidence \cite{blackburn2015}& Bayesian hierarchical models with efficient samplers, 
Gaussian mixture models \cite{delpozzo2018}
\\
\noalign{\vskip 1mm} 
\hline
\\
\end{tabu}
\end{center}
}

\end{sidewaystable}

\begin{sidewaystable}

\small{
    \centering
\leftline{Table 2: Astroinformatics and Astrostatistics Research in Thematic Areas - continued}
\begin{center}
\begin{tabu} to 1.02\textwidth{X[1,l]X[1,l]X[1,l]}
\hline
{\bf Advances} (issues under consideration)     & {\bf (Data/Analysis) Challenges}   &   {\bf Future } (Emergent Methods/Methodology) \\
 Methods/Methodology) \\
\hline
\noalign{\vskip 1.7mm}
\multicolumn{3}{l}{\bf Planetary Systems}  \\
\noalign{\vskip 1mm}
 $>$ 100,000 target stars identified in the Kepler 4-year mission \cite{jenkins2017, hsu2018}; characterization of planetary systems \cite{thompson2018}; analysis of transit timing variations to  characterize the exoplanet mass-radius relationship &
Inadequate statistical estimators used in early analysis \cite{dfm2014, hsu2018};
Bayesian hierarchical models to combine large measurement uncertainties and intrinsic astrophysical variability \cite{wolfgang2016,ning2018}; The combination of TESS and ground-based Doppler surveys is poised to significantly expand the sample of planets for such analyses, providing new constraints for physical modeling of exoplanets
& Advanced methods applicable to new generation of instruments; for characterizing stellar variability \cite{davis2017}; machine learning and data-driven models for analyzing high resolution spectroscopic time series \cite{jones2017,bedell2019}; quantifying the evidence of low-mass planets, even with simple RV time series \cite{dumusque2017, nelson2018}; Gaussian Process for RV time series \cite{czekala2017}
\\
\noalign{\vskip 1mm}
\hline
\noalign{\vskip 1mm}
\multicolumn{3}{l}{\bf Stars and Planet Formation} \\
\noalign{\vskip 1mm}
Extract information from the high resolution optical and infrared spectra; ALMA \cite{alma2015} images; use molecular lines to probe high-dimensional space with dynamic and chemical information 
multiwavelegth studies of spatial and kinematic structures in clusters\cite{Kuhn15}
& Gaussian processes to deal with correlated residuals from model systematics, and to construct physics-based forward-models \cite{czekala2015a, narayan2018}; data-driven approaches for 
accurate 
spectral models on a pixel-by-pixel basis; only over-simplified models are used \cite{czekala2015a} as complex models with chemical networks\cite{hogg2016}, non-ideal MHD,  effects of planet clearing, are computationally inefficient for Bayesian inference \cite{lyra2016, bai2017}; 
inadequate statistics for understanding the highly non-homogeneous distributions of young stellar objects 
& Bayesian inference for sophisticated models in a computationally tractable implementation; data-driven approaches to model disk rotation curves and search for the influence of planets \cite{yen2016,teague2018};
Fourier-based matched filters to enable sensitive and efficient molecular line searches (e.g., detection of faint molecular emission from methanol \cite{loomis2018}); 
new statistics for inhomogeneous point processes, modifications of the two-point correlation function \cite{Peebles80,Baddeley15};
advanced cluster analysis methods including mixture models and density clustering algorithms \cite{KuhnFeigelson17,Joncour18,Bovy11}\\
\noalign{\vskip 1mm}
%
\hline
\noalign{\vskip 1mm}
\multicolumn{3}{l}{\bf Cosmology and Fundamental Physics} \\
\noalign{\vskip 1mm}
Harness subtle signals to reduce scatter \cite{ntampaka2015, ntampaka2018, ho2019}, quickly generate mock data \cite{2018ComAC, He2018}
discriminate models to quantify statistical and systematic uncertainties 
\cite{deSouza2019a,deSouza2019b};
to perform likelihood-free cosmological inference \cite{Ishida2015, Hahn2017, Alsing2018, Kacprzak2018, Leclercq2018}
and classify objects \cite{Ishida2013,Varughese2015, Charnock2017, Dai2018, Lanusse2018, Ishida2019}
& Apply new advancements in ML interpretability, including saliency maps \cite{simoyan2013} and the deep k-nearest neighbors approach \cite{papernot2018}
& Development of a brand-new Bayesian machine learning framework suitable to handle the many layers of statistical error structures presented in Astronomy (selection bias, errors-in-measurements, systematics,\,etc.) \\
\noalign{\vskip 1mm}
\hline
\end{tabu}

\end{center}
}

\end{sidewaystable}

\pagebreak

\bibliographystyle{abbrv}

\end{document}